\documentclass[aps,preprint]{revtex4}%
\usepackage{amssymb}
\usepackage{amsfonts}
\usepackage{amsmath}
\usepackage{graphicx}%
\providecommand{\U}[1]{\protect\rule{.1in}{.1in}}
\begin{document}
\preprint{ }

\begin{center}
{\LARGE How Events Separated by a Timelike Interval\ Can Help Us Understand
Quantum Nonlocality}

\bigskip

{\large Luiz Carlos Ryff}

\textit{Instituto de F\'{\i}sica, Universidade Federal do Rio de Janeiro, Rio
de Janeiro, Brasil}

E-mail: ryff@if.ufrj.br\bigskip
\end{center}

{\small Quantum entanglement plays a fundamental role in quantum cryptography
and computation. An important example of quantum entanglement can be found in
the correlations of Einstein, Podolsky, and Rosen (EPR). However, despite the
plethora of articles related to the topic, different interpretations of the
EPR correlations coexist, and a consensus has not yet been reached. In this
article, we seek to demonstrate, through the simple and direct application of
quantum formalism, how events separated by timelike intervals can, strangely
enough, help us better understand some aspects of the so-called
\textquotedblleft quantum nonlocality\textquotedblright\ associated with EPR
correlations.}

\bigskip

\textbf{1. Introduction}

\bigskip

According to Einstein, Podolsky, and Rosen (EPR), Quantum Mechanics (QM) would
be an incomplete theory \textrm{[1]}. To demonstrate this, they used a
counterfactual argument and relied on the premise, based on the Special Theory
of Relativity (STR), that no interaction can propagate at a speed greater than
that of light in a vacuum \textrm{[2]}. In the EPR argument, we have two
particles distant from each other, but quantum mechanically entangled, which
enables us, by determining the position of the former, to know the position of
the latter. However, instead of determining the position, we could determine
the linear momentum of the former, which would allow us to know the linear
momentum of the latter. Since there could be no communication between the two
particles (we are considering events separated by a spacelike interval), this
would serve as evidence that the second particle would have well-defined
position and linear momentum. However, according to QM, it is not possible to
simultaneously\ determine the position and linear momentum of a particle with
arbitrary precision. In fact, according to Bohr, and in line with the concept
of wave-particle duality, a particle cannot \textit{have }well-defined
position and linear momentum at the same time. For Einstein -- a proponent of
determinism -- QM, which is essentially probabilistic, would be correct but incomplete.

The EPR argument was simplified by Bohm, who considered two particles in the
spin singlet state \textrm{[3]}. This modifies the argument on a crucial
point. When we make the first particle pass through a Stern-Gerlach (S-G)
apparatus (which can have an arbitrary orientation) and when it is detected,
it becomes clear that we are not, strictly speaking, measuring the orientation
of its spin; in reality, it is being \textquotedblleft
forced\textquotedblright\ to follow one of the two paths presented,
corresponding to \textquotedblleft spin up\textquotedblright\ and
\textquotedblleft spin down\textquotedblright. Therefore, it is surprising
that (according to QM, which we accept as correct) the second particle,
distant from the first, when passing through an S-G apparatus (which can also
have an arbitrary orientation) behaves as though having spin opposite to that
of the first, since, in principle, it would not have \textquotedblleft
information\textquotedblright\ about the orientation of the S-G apparatus
through which the first passed \textrm{[4]}. Naturally, to the extent that the
particle is \textit{forced}, the counterfactual reasoning plays a different
role here to that which it did in the original EPR argument. Instead of having
the freedom to choose between \textit{measuring} position or momentum, we have
the freedom to choose \textit{the orientation} of the S-G apparatus.

Excluding unlikely coincidences, our experience suggests to us, fundamentally,
two ways to attempt to explain the correlations between two particles that are
spatially distant from each other. The first (\textquotedblleft
local\textquotedblright\ interpretation) presupposes some common pre-existing
property to the particles, such that the particles have already been emitted
in correlated states, and that a measurement performed on any one of them
cannot have any effect on the other; the second (\textquotedblleft
nonlocal\textquotedblright\ interpretation) assumes some form of
\textquotedblleft communication\textquotedblright\ between them, such that a
measurement performed on either of them can influence the state of the other.
In the case of EPR correlations, Bell's theorem \textrm{[5]} can assist us in
evaluating which of the two explanations appears to be the correct one.
Experiments with entangled polarization photons \textrm{[6]},\ particularly
those that utilize two-channel polarizers (which play a role similar to that
of S-G apparatuses in the case of correlated spin particles), and those that
close possible loopholes, support QM (through violations of Bell's
inequalities) and, for many physicists, reinforce the second explanation
\textrm{[7] }(which would invalidate the basic premise of EPR). In the case of
pairs of photons with entangled polarizations, the polarization state in which
one of the photons is detected determines the polarization state in which the
other photon will be forced, even if the photons are moving away in opposite
directions and are distant from each other.\ This led Bell and Bohm to
conjecture that, in the case of EPR correlations, some physical influence
might be propagating at a speed greater than that of light, and that perhaps
the very theory of relativity needed to be modified \textrm{[8]}. On the other
hand, there are those who resist admitting any type of connection between
entangled particles, presenting alternative but equally controversial
interpretations, such as Many Worlds (MW), so-called Bayesian (Qbism, QB), and
others \textrm{[9]}. The indisputable fact is that the debate over whether QM
is a local or nonlocal theory continues to this day. Our intention in this
article is not to provide answers but to examine the correlations from a new
perspective and suggest questions that may, ultimately, represent a
contribution to the debate.

\bigskip

\textbf{2. EPR correlations in the case of events separated by a spacelike
interval}

\bigskip

In order to gain a better understanding of the subject and the difficulties of
interpretation that arise, we will examine a typical experiment involving
space-like events \textrm{(Fig. 1)}.\ A source $S$ emits pairs of photons
($\nu_{1}$, $\nu_{2}$) with entangled polarizations. The photon $\nu_{1}$%
($\nu_{2}$) reaches the two-channel polarizer I (II), where it can be
transmitted or reflected, and is then detected at D$_{1}$ (D$_{2}$) (if
transmitted) or D$_{1^{\prime}}$\ (D$_{2^{\prime}}$) (if
reflected).\textrm{\ }In the laboratory reference frame ($L_{1}$) the photons
simultaneously reach their respective polarizers (which are equally distant
from $S$) and detectors.%
\[%
%TCIMACRO{\FRAME{itbpFU}{3.5354in}{0.857in}{0in}{\Qcb{\QTR{small}{Fig. 1: EPR
%experiment in the case of events separated by a spacelike interval.}}}%
%{}{dois.eps}{\special{ language "Scientific Word";  type "GRAPHIC";
%maintain-aspect-ratio TRUE;  display "USEDEF";  valid_file "F";
%width 3.5354in;  height 0.857in;  depth 0in;  original-width 3.4886in;
%original-height 0.825in;  cropleft "0";  croptop "1";  cropright "1";
%cropbottom "0";  filename 'dois.eps';file-properties "XNPEU";}} }%
%BeginExpansion
{\parbox[b]{3.5354in}{\begin{center}
\includegraphics[
height=0.857in,
width=3.5354in
]%
{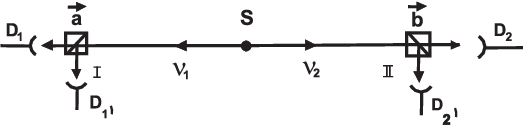}%
\\
{\protect\small Fig. 1: EPR experiment in the case of events separated by a
spacelike interval.}%
\end{center}}}
%EndExpansion
\]

\bigskip

Let us consider the entangled state in which, if the polarizers have the same
orientation (\textit{whatever it may be}), the photons will always be detected
with the same polarization:%
\begin{equation}
\left\vert \psi\right\rangle =\frac{1}{\sqrt{2}}(\left\vert x\right\rangle
_{1}\left\vert x\right\rangle _{2}+\left\vert y\right\rangle _{1}\left\vert
y\right\rangle _{2}), \tag{1}%
\end{equation}
where the kets $\left\vert x\right\rangle $ and $\left\vert y\right\rangle $
represent states with arbitrarily mutually orthogonal polarizations
\textrm{[10]}. We can observe the experiment from a Lorentz reference frame
($L_{2}$) in which $\nu_{1}$ is always detected first (a reference frame that
moves from right to left in \textrm{Fig. 1}, in which pol. I \textquotedblleft
approaches\textquotedblright\ $\nu_{1}$ and pol. II \textquotedblleft moves
away\textquotedblright\ from $\nu_{2}$). Knowing the polarization state in
which $\nu_{1}$ was detected, we can immediately \textit{infer} in which
polarization state $\nu_{2}$ will reach polarizer II, since the first
detection forces, so to speak, $\nu_{2}$ into a well-defined polarization
state (the same state in which $\nu_{1}$ was detected). Therefore, if the
polarizer is oriented in the $\mathbf{a}\ $direction,\ then $\nu_{2}$ must be
in a well-defined polarization state ($\left\vert a_{\parallel}\right\rangle
$, parallel to $\mathbf{a}$, or $\left\vert a_{\perp}\right\rangle $,
perpendicular to $\mathbf{a}$, depending on the channel in which $\nu_{1}$ was
detected) before reaching polarizer II. If polarizer II is oriented in the
$\mathbf{b}$ direction, we can establish, using Malus's law, the probability
of $\nu_{2}$ being transmitted or reflected, which gives us the probabilities
of coincident detections\textrm{: }%
\begin{equation}
p(a_{\parallel},b_{\parallel})=p(a_{\perp},b_{\perp})=\frac{1}{2}\cos^{2}(a,b)
\tag{2}%
\end{equation}
and
\begin{equation}
p(a_{\parallel},b_{\perp})=p(a_{\perp},b_{\parallel})=\frac{1}{2}\sin
^{2}(a,b), \tag{3}%
\end{equation}
where $p(a_{\parallel},b_{\perp})=p_{1}(a_{\parallel})p_{2}(b_{\perp}\mid
a_{\parallel})$, for example, is the probability of $\nu_{1}$ being
transmitted and $\nu_{2}$ being reflected, with $p_{1}(a_{\parallel})=1/2$
being the probability of $\nu_{1}$ being transmitted and $p_{2}(b_{\perp}\mid
a_{\parallel})=\sin^{2}(a,b)$ the probability of $\nu_{2}$ being reflected
when $\nu_{1}$ was transmitted, and so on \textrm{[11]. }

On the other hand, it is also possible to observe the experiment from a
Lorentz reference frame ($L_{3}$) in which $\nu_{2}$ is always detected first
(a reference frame that moves from left to right in \textrm{Fig. 1}, where
pol. II \textquotedblleft approaches\textquotedblright\ $\nu_{2}$ and pol. I
\textquotedblleft moves away\textquotedblright\ from $\nu_{1}$), thereby
forcing $\nu_{1}$ into a well-defined polarization state. Thus, for an
\textquotedblleft observer\textquotedblright\ in $L_{3}$, it is $\nu_{1}$ that
must be in a well-defined polarization state before reaching polarizer I,
whereas $\nu_{2}$ will not be in any defined polarization state before
reaching polarizer II. In this case, we would have, for example,\textrm{\ }
$p(a_{\parallel},b_{\perp})=p_{2}(b_{\perp})p_{1}(a_{\parallel}\mid b_{\perp
})$, where $p_{2}(b_{\perp})=1/2$ is the probability of\textrm{\ }$\nu_{2}$
being reflected and\textrm{\ }$p_{1}(a_{\parallel}\mid b_{\perp})=\sin
^{2}(a,b)$ is the probability of\textrm{\ }$\nu_{1}$ being transmitted
when\textrm{\ }$\nu_{2}$ has been reflected, and so on and so forth.

Naturally, regardless of the reference frame used to describe the experiment,
$L_{1}$, $L_{2}$ or $L_{3}$, the probabilities will always be given by
equations (2) and (3). We see that the descriptions (not the predictions) in
$L_{2}$ and $L_{3}$ are mutually contradictory, which seems to undermine the
idea that there could be any interaction propagating from one photon to the
other (would it be from $\nu_{1}$ to $\nu_{2}$ or from $\nu_{2}$ to $\nu_{1}%
$?), despite the observed correlations that violate Bell's inequalities
\textrm{[12]}, and the very idea of attributing an external and objective
reality to the polarization of the second photon of the entangled pair
\textit{before} it reaches the polarizer seems to make no sense \textrm{[13]
}(by \textquotedblleft objective\textquotedblright\ is meant \textquotedblleft
independent of the observer\textquotedblright). Naturally, observers in
different Lorentz frames may have conflicting descriptions regarding the
temporal order of events separated by a spacelike interval. This only becomes
a problem if we admit some kind of causal relation between the events, as
suggested by Bell and Bohm in the case of EPR correlations.

\bigskip\ 

\textbf{3. EPR correlations in the case of events separated by a timelike
interval}

\bigskip

However, there is another way to approach EPR correlations, which suggests a
different conclusion. To do this, we must consider timelike events and the
following thought experiment. In our experiment with polarization-entangled
photons, we will now introduce a deviation in the path of $\nu_{2}$ (M =
mirror), so that $\nu_{1}$ is always detected first in any Lorentz reference
frame \textrm{(Fig. 2)}.%
\[%
%TCIMACRO{\FRAME{itbpFU}{4.4097in}{1.7616in}{0in}{\Qcb{\QTR{small}{Fig. 2: EPR
%experiment in the case of events separated by a timelike interval.}}}%
%{}{um.eps}{\special{ language "Scientific Word";  type "GRAPHIC";
%maintain-aspect-ratio TRUE;  display "USEDEF";  valid_file "F";
%width 4.4097in;  height 1.7616in;  depth 0in;  original-width 4.3587in;
%original-height 1.7236in;  cropleft "0";  croptop "1";  cropright "1";
%cropbottom "0";  filename 'um.eps';file-properties "XNPEU";}} }%
%BeginExpansion
{\parbox[b]{4.4097in}{\begin{center}
\includegraphics[
height=1.7616in,
width=4.4097in
]%
{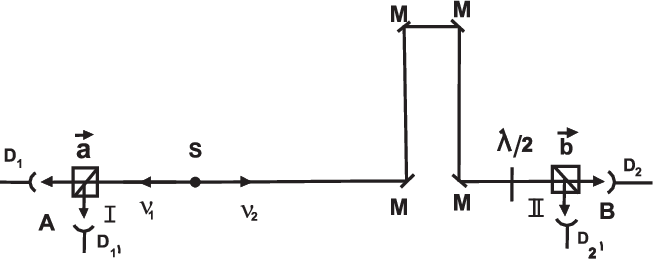}%
\\
{\protect\small Fig. 2: EPR experiment in the case of events separated by a
timelike interval.}%
\end{center}}}
%EndExpansion
\]

\bigskip\ 

Therefore, Alice (A, near polarizer I) , knowing the polarization state in
which $\nu_{1}$ was detected, could send a message that was received by Bob
(B, near polarizer II) before $\nu_{2}$ reached polarizer II, allowing B to be
aware of the polarization state of $\nu_{2}$ before it impinges on polarizer
II. B could then confirm the information provided by A, either by rotating his
polarizer to different orientations and confirming Malus's law, or by using a
half-wave plate ($\lambda/2$) to change the polarization of $\nu_{2}$. For
instance, B can change the state of $\nu_{2}$ from $\left\vert a_{\parallel
}\right\rangle $ to $\left\vert b_{\parallel}\right\rangle $ whenever $\nu
_{1}$ is transmitted, and from $\left\vert a_{\perp}\right\rangle $\ to
$\left\vert b_{\parallel}\right\rangle $ \ whenever $\nu_{1}$ is reflected,
ensuring that $\nu_{2}$ is always detected in the transmission channel. This
allows B to ascertain that the possible states of $\nu_{2}$ would depend on
the orientation chosen by A for polarizer I; they will be $\left\vert
a_{\parallel}\right\rangle $ and $\left\vert a_{\perp}\right\rangle $, if the
orientation was $\mathbf{a}$, $\left\vert c_{\parallel}\right\rangle $ and
$\left\vert c_{\perp}\right\rangle $, if the orientation was $\mathbf{c}$, and
so forth. This would eliminate any possibility of explaining the correlations
as resulting from properties previously shared by the particles (we are not
taking into account hypothetical `conspiracies' of nature that could `mimic'
the results predicted by QM, while simultaneously preserving locality, which
have already been largely ruled out through loophole-free tests of Bell's
inequalities \textrm{[6]}).

In this manner, we can posit an argument (in line with the position defended
by Bell and Bohm \textrm{[8]}) in favor of the viewpoint that what happens to
the first photon has an influence on the state of the second, changing it from
an undefined to a defined polarization state. It seems reasonable to conclude
that, unlike what happened in the case of spacelike events, it is possible to
ascribe an external and objective reality to the polarization of the second
photon after the detection of the first (in the sense of being independent of
the frame of reference used to describe the experiment). Undoubtedly, it is
possible to conjecture that, in the case of timelike events, the detection of
the first photon, after it passes through the polarizer, forces the second
into a defined polarization state through some unknown interaction propagating
at a speed equal to or less than that of light. However, the mere fact that we
acknowledge this possibility may have profound consequences for our
understanding of EPR correlations. For instance, we can reduce the height of
the deviation in the path of $\nu_{2}$ to have spacelike events, in an attempt
to verify whether the correlations would disappear, which should happen if
they resulted from this supposed unknown interaction. Nevertheless the
correlations should remain, as already experimentally verified \textrm{[6]},
which would seem to suggest an exchange of information between
quantum-entangled particles, even in the case of spacelike events
\textrm{[7]}. Obviously, correlations do not necessarily imply
cause-and-effect relationships, and the fact that events are separated by
timelike intervals does not imply a causal relationship between them. What is
noteworthy in this case is the fact that the orientation of pol. I determines
the states in which $\nu_{2}$ will reach pol. II. Naturally, we should not
conclude that we have a definitive argument in favor of the point of view
advocated by Bell and Bohm, but we certainly have an argument that would be in
line with the hypothesis they raised. However, as we will see below, trying to
reconcile this point of view with STR is far from an easy task, if it is even feasible.

\bigskip

\textbf{4. Discussion}

\bigskip

Naturally, if we are to accept that in the experiment represented in
\textrm{Fig. 2}, the detection of the first photon ($\nu_{1}$) causes the
second ($\nu_{2}$) to transition from an undefined to a well-defined
polarization state, the following question arises: from what moment (which
will depend on the Lorentz reference frame used to describe the experiment)
can we assign a well-defined polarization to the second photon? Or more
specifically: the transition from entangled $\nu_{2}$ and, therefore, without
a well-defined polarization, to $\nu_{2}$ with well-defined polarization would
not be an objective fact (namely, independent of the observer), in the sense
that coincident events in space and time should be seen as coincident in any
Lorentz reference frame? In our case, the coincident events would be the
transition from the undefined polarization state to the defined polarization
state and the localization of the photon packet at the moment this transition
occurs.\ However, assuming that the detection of $\nu_{1}$ instantaneously
forces $\nu_{2}$ into a well-defined state of polarization, in the laboratory
reference frame ($L_{1}$) the transition occurs before $\nu_{2}$ reaches the
first mirror ($M$) (the distance between $S$ and $D_{1} $ and $S$ and
$D_{1^{\prime}}$ is less than the distance between $S$ and $M$), whereas in a
reference frame that is moving from left to right ($L_{3}$), for example, the
transition can occur, by appropriately choosing the speed of $L_{3}$, after
$\nu_{2}$ has passed from $M$, since $\nu_{1}$ takes longer to be detected
(polarizer I and detectors $D_{1}$ and $D_{1^{\prime}} $ are \textquotedblleft
moving away\textquotedblright\ from $\nu_{1}$, while the mirror $M$ is
\textquotedblleft approaching\textquotedblright\ $\nu_{2}$). It is easy to
verify that the velocity of\textrm{\ }$L_{3}$ must satisfy the condition
$v>c(x_{2}-x_{1})/(x_{2}+x_{1})$, where\textrm{\ }$x_{1}$ is the distance
between $S$ and\textrm{\ }$D_{1}$, and $x_{2}\ $is the distance between $S$
and\textrm{\ }$M$ (where $x_{1}$ and $x_{2}$ may refer to the distances
measured in $L_{1}$ or $L_{3}$).

It is also interesting to note that before $\nu_{2}$\ reaches the first
mirror, it is not yet decided which experiment is involved, whether it is the
one in Fig. 1 or the one in Fig. 2, which makes it problematic to know when it
is actually possible to assign a well-defined polarization to $\nu_{2}$. For
example, if the two lower mirrors ($M$) in \textrm{Fig. 2} were to be replaced
by beam splitters, whenever $\nu_{2}$ is reflected (transmitted) the
detections of $\nu_{1}$ and $\nu_{2}$ will be events separated by a timelike
(spacelike) interval. For the observer in $L_{1}$, whenever $\nu_{2}$ takes
the long path, it is possible to verify that the detection of $\nu_{1}$ forced
$\nu_{2}$ into a well-defined polarization state, since the information sent
by A reaches B before $\nu_{2}$. The same applies to an observer in $L_{3}$.
On the other hand, whenever $\nu_{2}$ takes the short path, the observer in
$L_{1}$ can infer, based on the previous case, that the detection of $\nu_{1}$
forced $\nu_{2}$ into a well-defined polarization state, the only difference
being that now the information arrived late, namely after $\nu_{2}$. However,
for the observer in $L_{3}$, now is $\nu_{2}$ that is detected first, forcing
$\nu_{1}$ into a well-defined polarization state.

\bigskip

\textbf{5. Conclusion}

\bigskip

The scope of this article is to present EPR correlations from a new
perspective, in order to gain a more \ `tangible' understanding of the
subject. Our intention is to try to demonstrate how the simple and direct
application of QM allows us to highlight some interesting and counterintuitive
aspects related to EPR correlations. We saw that, through timelike events, it
would be possible to ascertain, by observing what happens with the second
photon ($\nu_{2}$), that the orientation of polarizer I, on which the first
photon ($\nu_{1}$) impinges, determines the states in which $\nu_{2}$ can be
found before reaching polarizer II, which suggests that the observed
correlations cannot be a consequence of common pre-existing properties of the
particles. Interestingly, it is not possible to determine with certainty at
what moment the second photon transitions from an undefined to a defined
polarization state. This raises a few interesting questions. What kind of
interaction, if any, might be responsible for this correlation between
entangled photons \textrm{[14]}? Might they constitute a single entity, even
being spatially distant from one another \textrm{[15]}? Should we acknowledge,
considering spacelike events, that behind the scenes something is moving
faster than light, as conjectured by Bell and Bohm? As demonstrated, if the
EPR correlations result from causal influences propagating at a finite
superluminal speed, then the possibility of superluminal communication cannot
be dismissed \textrm{[16]}. Might it be possible to reconcile this result with
the STR? Might it be feasible to introduce a privileged reference frame, which
could prevent causal paradoxes \textrm{[8]}? Might Lorentz symmetry be broken
in the case of quantum nonlocality, and the equivalence between active and
passive transformations not be valid \textrm{[17]}? These are questions that
deserve our reflection and seek to escape the tendency of \textquotedblleft
calculating without asking\textquotedblright. As emphasized by John Bell:
\textquotedblleft The scientific attitude is that correlations cry out for
explanation\textquotedblright\ \textrm{[18]}.

In closing: It was not our goal in this article to analyze different
interpretations, much less to defend any particular viewpoint; our intention
was simply to show how the analysis of events separated by a timelike interval
can assist in making the difficult conceptual questions raised by EPR
correlations more evident.


\begin{thebibliography}{99}
\bibitem {4}Might also be possible to assume, in the original EPR argument,
that when we perform a measurement we are, in fact, \textquotedblleft
forcing\textquotedblright\ the particle to \textquotedblleft
choose\textquotedblright\ a position, or to \textquotedblleft
select\textquotedblright\ a momentum; however, this point is less evident in
this case, and it does not seem to us that this idea was present in the
original argument, which is fundamentally based on the premise, supported by
classical physics, that a particle must simultaneously possess a well-defined
position and linear momentum.
\bibitem {13}We are referring to the polarization of the second photon after
the first ($\nu_{1}$ in $L_{2}$, $\nu_{2}$ in $L_{3}$) has been detected.
Naturally, before any measurement, or to be more precise, before the first
photon is detected in a well-defined polarization state, the entangled state
is represented by Eq. (1), and no polarization can be associated with either
of the photons. For an observer in $L_{1}$, there is no first photon to be
detected, since we are assuming that, in principle, the detections are
coincident in this case. In fact, due to inevitable \ practical limitations,
it is always possible to assume that one of the photons was detected (or
annihilated) first, thereby forcing the other into a well-defined polarization state.
\bibitem {15}A rather peculiar single entity, in fact, since when one of the
photons from an entangled pair passes through an optical device, nothing
happens to the other. Only when a detection occurs do they seem to behave as a
single entity.
\end{thebibliography}
\end{document}